\documentstyle[11pt]{article}

\newlength{\minitwocolumn}
\catcode`\@=11
\def\relaxnext@{\let\next\relax}
\font\tenmsy=msym10 scaled\magstep1
\font\sevenmsy=msym7 scaled\magstep1
\font\fivemsy=msym5  scaled\magstep1
\newfam\msyfam
\textfont\msyfam=\tenmsy
\scriptfont\msyfam=\sevenmsy
\scriptscriptfont\msyfam=\fivemsy
\font\teneuf=eufm10 scaled\magstep1
\font\seveneuf=eufm7 scaled\magstep1
\font\fiveeuf=eufm5 scaled\magstep1
\newfam\euffam
\textfont\euffam=\teneuf
\scriptfont\euffam=\seveneuf
\scriptscriptfont\euffam=\fiveeuf
\def\frak{\relaxnext@\ifmmode\let\next\frak@\else
 \def\next{\Err@{Use \string\frak\space only in math mode}}\fi\next}
\def\goth{\relaxnext@\ifmmode\let\next\frak@\else
 \def\next{\Err@{Use \string\goth\space only in math mode}}\fi\next}
\def\frak@#1{{\frak@@{#1}}}
\def\frak@@#1{\noaccents@\fam\euffam#1}
\def\Bbb{\relaxnext@\ifmmode\let\next\Bbb@\else
 \def\next{\Err@{Use \string\Bbb\space only in math mode}}\fi\next}
\def\Bbb@#1{{\Bbb@@{#1}}}
\def\Bbb@@#1{\noaccents@\fam\msyfam#1}
\def\accentfam@{7}
\def\noaccents@{\def\accentfam@{0}}
\catcode`\@=\active
\newcommand{\bz}{{\Bbb Z}}

\newcommand{\bc}{{\Bbb C}}

\def\Remark{\bigskip\noindent{\sl Remark.}\quad}

\def\eq#1\endeq{\begin{eqnarray}#1\end{eqnarray}}
\def\eqn#1\endeqn{\begin{eqnarray*}#1\end{eqnarray*}}

\def\e{\varepsilon}

\def\d{\delta}
\def\End{{\rm End}\,}

\makeatletter
\@addtoreset{equation}{section}
\makeatother

\newtheorem{thm}{Theorem}[section]
\newtheorem{prop}[thm]{Proposition}
\newtheorem{lem}[thm]{Lemma}

\begin{document}
\begin{flushright}
RIMS-945 \\
Sept. 1993
\end{flushright}
\vspace{24pt}
\begin{center}
\begin{Large}
{\bf Smirnov's Integrals and \\
Quantum Knizhnik-Zamolodchikov
Equation of Level $0$}
\end{Large}

\vspace{36pt}
Michio Jimbo\raisebox{2mm}{{\scriptsize 1}},
Takeo Kojima\raisebox{2mm}{{\scriptsize 2}},
Tetsuji Miwa\raisebox{2mm}{{\scriptsize 2}}
and
Yas-Hiro Quano\raisebox{2mm}{{\scriptsize 2}$\star$}

\vspace{6pt}

{}~\raisebox{2mm}{{\scriptsize 1}}
{\it  Department of Mathematics, Faculty of Science,
      Kyoto University, Kyoto 606, Japan}

{}~\raisebox{2mm}{{\scriptsize 2}}
{\it Research Institute for Mathematical Sciences,
     Kyoto University, Kyoto 606, Japan}

\vspace{72pt}

\underline{Abstract}

\end{center}

We study the quantum Knizhnik-Zamolodchikov equation
of level $0$ associated with
the spin $1/2$ representation of
$U_q \bigl(\widehat{\frak s \frak l _{2}}\bigr)$.
We find an integral formula for solutions
in the case of an arbitrary total spin and $|q|<1$.
In the formula, different solutions can be
obtained by taking different integral kernels
with the cycle of integration being fixed.

\vspace{24pt}

\vfill
\hrule

\vskip 3mm
\begin{footnotesize}

\noindent\raisebox{2mm}{$\star$}A Fellow of
the Japan Society for the Promotion of Science for
Japanese Junior Scientists.
\end{footnotesize}
\newpage

\section{Introduction}

In this paper, following \cite{Smbk},
we will give an integral formula for solutions to
the quantum Knizhnik-Zamolodchikov (KZ) equation \cite{FR}
for the quantum affine algebra
$U_q\bigl({\widehat{\frak s \frak l}}_{2}\bigr)$
when the spin is $1/2$, the level is $0$ and $|q|<1$.

In \cite{Smbk}, Smirnov gave an integral formula for the form-factors
of the sine-Gordon model. His method was
solving a system of difference equations for a vector-valued function
in $N$ variables $(\beta_1,\ldots,\beta_N)$
which takes values in the $N$-fold tensor product of the spin $1/2$
representation ${\bc}^2$ of $U_q\bigl({\widehat{\frak s \frak l}}_{2}\bigr)$.
The total space ${\bc}^2\otimes\cdots\otimes{\bc}^2$
splits into the subspaces of fixed total spins $(l-n)/2$
where $l+n=N$ and $0\leq n \leq N$.
In \cite{Smbk}, an integral formula was given
for the case $n=l=N/2$ ($N$:even) and $|q|=1$.
In a private communication to the present authors,
Smirnov showed the modified formula for the case $|q|<1$, which is given
at the end of Section 4 of the present paper.
Our main contribution is to generalize Smirnov's formula to
the case of an arbitrary total spin.

Before going into the details,
let us discuss several points on the
quantum Knizhnik-Zamolodchikov
equation and the integral formulas.
We are largely indebted to Smirnov
for discussions on this matter.

The KZ equation was introduced
in \cite{KZ} as the master equation
for the correlation functions
of the conformal field theories with
gauge symmetries, i.e.,
the Wess-Zumino-Witten model.
In \cite{TK}, it was studied
by using the representation theory of the
affine Lie algebra
${\widehat{\frak s \frak l}}_{2}$.
The main observation in those
developments is that the local operators
in the Wess-Zumino-Witten model
is realized as the intertwiners of the highest weight
representations
(they are called the vertex operators),
and the correlation functions,
that are given as the matrix elements
of the products of the vertex operators
with respect to the highest weight vectors,
satisfy the KZ equation.
The components of the intertwiners belong to
finite-dimensional representations  of the
affine Lie algebras,
e.g., the spin $1/2$ representation ${\bc}^2$ of
${\widehat{\frak s \frak l}}_{2}$.
Such representations
are parameterized by the complex spectral
parameters.
The correlation functions,
therefore, take their values
in the tensor products of
finite-dimensional representations and
depend on the spectral parameters associated with them,
say $(\beta_1,\ldots,\beta_N)$.

This similarity of the KZ equation
to the form-factor equation led to the
introduction of the quantum KZ equation
by Frenkel and Reshetikhin.
The representation theory of the affine
Lie algebras is replaced
by that of the quantum affine algebras, and
the system of differential equations
is replaced by a system of
difference equations.

Compared with the quantum KZ equation
of Frenkel and Reshetikhin,
Smirnov's equation for the sine-Gordon model
is special in the following
two points.
The former contains a complex parameter,
the level of
the highest weight representations.
In the latter, the level is set to
$0$. This is the first point.
In this paper, we stick to the level $0$ case
as in Smirnov's work.

The second point is
that the quantum KZ equation admits a choice of a diagonal operator,
which acts on the tensor component whose
spectral parameter undergoes a shift (see (\ref{qKZ})).
Smirnov's equation chooses the
identity operator for that.
In our integral formula,
the said operator is not the identity in general,
but is fine-tuned according to
the total spin $(l-n)/2$.
Since Smirnov considered
only the case $l=n$, the fine tuning was
trivial.
To be precise,
the case $l=n\pm2$ is also studied in \cite{Smbk}.
However, since it is only in the rational limit,
the operator again reduces to the identity.

We now discuss the solutions and the integral formulas.

The analytic structures of the solutions
are different between
the quantum KZ equation
and the ordinary one.
The former admits solutions meromorphic
 in the variables
$(\beta_1,\ldots,\beta_N)$, the additive
spectral parameters.
The latter, in general, forces its solutions to exhibit
branches.
Namely, the monodromy structures
of the solutions are very different.
In fact, it is more important
to consider the braid relations for the solutions,
i.e., the behavior of the solutions
when two of the variables
are interchanged.
In \cite{TK}, the braid representations
for the KZ equation
was nicely deduced
from the commutation relation for the vertex operators.
As in \cite{FR,XXZ},
the quantum vertex operators enjoy a similar
commutation relation.
The difference in these two commutation relations is
that in the former the coefficients are independent of the
spectral parameters but they are not in the latter.

The said difference is closely related
to the following fact.
Being a
holonomic system of differential equations,
the space of the solutions to the
KZ equation is finite-dimensional,
while the quantum KZ equation admits
the linearity with the coefficients in the ring of
quasi-constants,
i.e.,  functions invariant under the relevant shifts of
the spectral parameters.
Therefore, the braiding property of the
solutions to the KZ equation
is unique up to finite-dimensional
similarity transformations,
while it is quite ambiguous for the quantum
KZ equation.
Nevertheless,
the commutation relations of the quantum
vertex operators imply
that the braid relation with the $R$-matrix
in the coefficients
(we call it the $R$-matrix symmetry, see (\ref{eqn:R-symm})),
are indeed compatible with the difference equation. E.g.,
the solutions obtained from the quantum vertex operators
satisfy that property.
Smirnov suggests that we can utilize the freedom of the
choice of the quasi-constants
to choose this particular braid relation
for the solutions.
Thus, we demand the $R$-matrix symmetry in constructing the
integral formulas.
Then, the vector-valued unknown function
reduces to a single function,
and the system of the difference equations
reduces to a certain
deformed cyclicity (\ref{eqn:cyc}) for it.

Once the equation is thus reduced,
we can construct the integral formula
as follows.
Let us explain our formula
in the case $n<l$.
We find it convenient to employ the multiplicative
spectral parameters $(z_1,\ldots,z_N)$,
where $\beta_i\propto\hbox{\rm log\/}z_i$.
Following Smirnov,
we choose an integral kernel
$\Psi(x_1,\ldots,x_n|z_1,\ldots,z_N)$.
It has simple poles at
$x_{\mu }=z_j q^{\pm(1+4k)}\quad(k=0,1,\cdots)$,
and satisfy
certain quasi-periodicities
with respect to the shifts of the
variables $x_{\mu }\rightarrow x_{\mu }q^4 $
and $z_j\rightarrow z_j q^4 $.
In fact, the choice of such an integral kernel
is not unique, because
the solutions to the quasi-periodicity conditions are not unique.
Smirnov's idea is that
this freedom in the integrand corresponds to
the freedom of the solutions up to quasi-constants.
The total integrand
contains a fixed rational function
$F=F(x_1,\ldots,x_n|z_1,\ldots,z_N)$
other than the said kernel.
It is a polynomial in $x_{\mu }$'s
and homogeneous in $x_{\mu }$'s and $z_j$'s.
We consider the total integral
as an integral transform of this rational
function in terms of the integral kernel,
fixing appropriately the cycles for
the integrations with respect to $x_{\mu }$'s.
It is easily shown that if
$F$ is a polynomial of degree greater than $N-1$
in a variable $x_{\mu }$, then
it is reducible to a lower degree polynomial
modulo a certain `total difference' which
vanishes after integration with respect to  $x_{\mu }$.
Using this fact we can write down a system of
algebraic relations for $F$ as a sufficient condition for the
deformed cyclicity.
These conditions are also necessary if we
assume that there is no other lower degree relations
of the kind mentioned above.
Finally, we find $F$ satisfying them.
The details will be given in the subsequent sections.

Lastly, we discuss the related works
on the integral formulas for
the quantum KZ equations.
In \cite{Matsuo,Reshet} solutions by
Jackson-type integrals are obtained.
Their formulas are in principle valid for general level,
as opposed to our integral formula restricted to level $0$.
On the other hand, the problem of choosing the cycles for Jackson-type
integrals, which accommodates the freedom of the solutions,
is not well studied.
In particular,
the choice of the cycles that leads to the $R$-matrix symmetry
is totally unclear.

A different type of integral formula
was obtained in \cite{CORR} by using the Frenkel-Jing
bosonization of the level $1$ highest weight representations \cite{FJ}.
Though the level can be chosen arbitrarily,
it gives only one particular solution for each level.
Since the relation between the formulas
in this paper and therein are not yet clear,
we do not discuss further on this matter.

The rest of the paper is organized as follows.
In Section 2, we formulate
the difference equation
in terms of a single component of the vector.
In Section 3, we solve the algebraic relation
for $F$ in the case
$n=1$. In Section 4 we solve the general
case recursively.

{}~

\section{Difference Equations}
The purpose of this section is
to formulate the problem, thereby
fixing our notations.

Let us begin by recalling the standard trigonometric
$R$-matrix $R(z)\in \End(V\otimes V)$
associated with $V=\bc ^2$.
Fix a complex number $q$ such that $0<|q|<1$.
The matrix $R(z)$ is specified by giving the
matrix elements relative
to the standard basis $v_+,v_-\in V$:
$$
R(z)v_{\e'_1}\otimes v_{\e'_2}=
\sum_{\e_1,\e_2}v_{\e_1}\otimes v_{\e_2}
R(z)^{\e_1\e_2}_{\e'_1\e'_2}.
$$
The nonzero entries are
\eqn
&&R(z)^{++}_{++}=R(z)^{--}_{--}=a(z),\\
&&R(z)^{+-}_{+-}=R(z)^{-+}_{-+}=b(z),\\
&&R(z)^{+-}_{-+}=R(z)^{-+}_{+-}=c(z),
\endeqn
where
\eqn
a(z)=1,\quad
b(z)=
{(1-z)q\over 1-q^2z},\quad
c(z)=
{(1-q^2)\sqrt{z}\over 1-q^2 z}.
\endeqn
In what follows we shall work with
the tensor product $V^{\otimes  N}$.
Following the usual convention
we let $R_{jk}(z)$ ($j\neq k$)
signify the operator on
$V^{\otimes N}$ acting as $R(z)$
on the $(j,k)$-th tensor
components and as identity
on the other components.
In particular we have
$R_{kj}(z)=P_{jk}R_{jk}(z)P_{jk}$, where
$P\in\End(V\otimes V)$
stands for the transposition
$P(x\otimes y)=y\otimes x$.

The main properties of $R(z)$
are the Yang-Baxter equation
\eq
R_{12}(z_1/z_2)
R_{13}(z_1/z_3)
R_{23}(z_2/z_3)
=R_{23}(z_2/z_3)
R_{13}(z_1/z_3)
R_{12}(z_1/z_2)
\label{eqn:YBE}
\endeq
and the unitarity relation
\eq
R_{12}(z_1/z_2)R_{21}(z_2/z_1)=1.
\label{eqn:unit}
\endeq

The equations we are concerned with
in this paper are the following
ones for a function $G(z_1,\cdots,z_N)$
with values in $V^{\otimes  N}$:

\noindent{\bf 1. $R$-matrix symmetry}
\begin{equation}
P_{j\,j+1} G (\cdots,z_{j+1},z_j,\cdots)
\quad =
R_{j\,j+1}(z_j/z_{j+1})G (\cdots,z_j,z_{j+1},\cdots)
\qquad (1\le j\le N-1),
\label{eqn:R-symm} \\
\end{equation}

\noindent{\bf 2. Deformed Cyclicity}
\begin{equation}
P_{12}\cdots P_{N-1 N} G (z_2,\cdots,z_N, z_1 q^{-4})
=
D_1 G (z_1,\cdots,z_N).
\label{eqn:cyc}
\end{equation}
In (\ref{eqn:cyc})
$D_1$
is an operator acting on the first component as
$D={\rm diag}(\delta _+ , \delta _- )$,
whose entries will be specified below,
and as identity on the other ones.
Here a remark is in order about the precise meaning of these equations.
Throughout this article the functions we consider
are not necessarily single valued in
$z_j$ but are meromorphic in the variable
$\log z_j$.
Accordingly the shift
$z_j\rightarrow z_jq^{-4}$
as in (\ref{eqn:cyc}) is understood to mean
$\log z_j\rightarrow \log z_j-4\log q$.

The equations (\ref{eqn:R-symm}) and
(\ref{eqn:cyc}) with $D=1$
appeared in  Smirnov's
works on the form factors of massive
integrable field theories \cite{Smbk}.
As was pointed out in \cite{Sm1} they imply the quantum
Knizhnik-Zamolodchikov equation of level $0$ \cite{FR}
\begin{equation}
\begin{array}{l}
G (z_1,\cdots,z_jq^4 ,\cdots,z_N)
=
R_{j-1\,j}(z_{j-1}/z_j q^4 )^{-1}
\cdots
R_{1\,j}(z_{1}/z_j q^4 )^{-1}D_j^{-1} \\
\qquad\qquad \times
R_{j\,N}(z_{j}/z_N)
\cdots
R_{j\,j+1}(z_{j}/z_{j+1})
G (z_1,\cdots,z_j,\cdots,z_N).
\label{qKZ}
\end{array}
\end{equation}
These equations have a $\bz _2$-symmetry which means that
if $G(z_1,\cdots,z_N)$ solves
(\ref{eqn:R-symm}--\ref{eqn:cyc}), then
$\tilde{G}(z_1,\cdots,z_N)=
\sigma^x\otimes\cdots\otimes\sigma^x\,
G(z_1,\cdots,z_N)$ with
$\sigma^x=\left(\matrix{0&1\cr 1&0\cr}\right)$
solves the same system
wherein $\d_+$ and $\d_-$ are interchanged.

In the sequel we set $\tau=q^{-1}$.
Define the components of $G$ by
\begin{eqnarray}
&&G(z _1 , \cdots , z _N ) =
\sum_{\varepsilon _j =\pm }
v_{\varepsilon _1 } \otimes \cdots
\otimes v_{\varepsilon _N }
\bigl(\prod_{\e_j <0} \sqrt{z}_j\bigr)\,
G^{\varepsilon _1 \cdots \varepsilon _N }
(z _1 , \cdots , z_N ). \label{eqn:components}
\end{eqnarray}
Then the equations (\ref{eqn:R-symm}) and (\ref{eqn:cyc})
read respectively as
\begin{eqnarray}
G ^{\cdots \stackrel{j}{\e} \stackrel{j+1}{\e} \cdots}
(\cdots,z_j,z_{j+1},\cdots)
&=&
G ^{\cdots \stackrel{j}{\e}\stackrel{j+1}{\e} \cdots}
(\cdots,z_{j+1},z_j,\cdots),
\label{eqn:Rsym1} \\
G ^{\cdots \stackrel{j}{+} \stackrel{j+1}{-} \cdots}
(\cdots,z_j,z_{j+1},\cdots)
&=&
{z_j-z_{j+1}\tau^2 \over (z_j-z_{j+1})\tau }
G ^{\cdots \stackrel{j}{-} \stackrel{j+1}{+} \cdots}
(\cdots,z_{j+1},z_j,\cdots) \nonumber \\
&-&
{(1-\tau^2)z_j \over (z_j-z_{j+1})\tau}
G^{\cdots \stackrel{j}{-} \stackrel{j+1}{+} \cdots}
(\cdots,z_j,z_{j+1},\cdots),
\label{eqn:Rsym} \\
G ^{\e_2 \cdots \e_N \e_1}(z_2,\cdots,z_N,z_1 \tau^4 )
&=&
\d_{\e_1}~
G ^{\e_1\e_2 \cdots \e_N} (z_1,z_2,\cdots,z_N).
\label{eqn:cyclic}
\end{eqnarray}
The factor  $\prod_{\e_j <0} \sqrt{z}_j$
in (\ref{eqn:components}) is so chosen that the coefficients in
in these equations are free from square root symbols.
Note that the singularity at $z_j =z_{j+1}$
in (\ref{eqn:Rsym}) is spurious.
The equations (\ref{eqn:Rsym1}--\ref{eqn:cyclic})
split into blocks, each involving components such that
$$
n=\sharp\{j\mid \e_j=-\}, \quad l
=\sharp\{j\mid \e_j=+\} \qquad(n+l=N)
$$
are fixed.
Because of the $\bz _2$-symmetry we may
assume $n\le l$ without loss of generality.

Consider the extreme component
\eq
G^{\overbrace{-\cdots-}^n\overbrace{+\cdots +}^l}
(z_1,\cdots,z_N)
=H(z_1,\cdots,z_n\,|\,z_{n+1},\cdots,z_N).
\label{eqn:1_comp}
\endeq
Because of (\ref{eqn:Rsym1})
this function is symmetric
separately in the variables
$(z_1,\cdots,z_n)$ and $(z_{n+1},\cdots,z_N)$.
The equation (\ref{eqn:Rsym}) tells that
all the components with fixed $n,l$
are uniquely determined from $H$.
Conversely given any such $H$
the Yang-Baxter equation guarantees
that (\ref{eqn:R-symm}) can be
solved consistently
under the condition (\ref{eqn:1_comp}).

\Remark An explicit way of
reconstructing $G$ from $H$ is described
in \cite{Smbk}.
The procedure goes as follows.
Define the operator $B(z_1,\cdots,z_N| t)\in\End(V^{\otimes N})$  by
\[
R_{1\,N+1}(z_1/t)\cdots R_{N\,N+1}(z_N/t)
=
\left(\matrix{ A(z_1,\cdots,z_N| t) & B(z_1,\cdots,z_N| t) \cr
              C(z_1,\cdots,z_N| t) & D(z_1,\cdots,z_N| t) \cr}\right).
\]
Here the $2\times 2$ matrix structure is
defined relative to the base
$v_\pm$ of
the $(N+1)$-th tensor component
of $V^{\otimes (N+1)}$.
For $\alpha=(\alpha_1,\cdots,\alpha_N)$
with $\alpha_i=\pm$, set
$J^{\alpha}_\pm=\{~j~|~\alpha_j=\pm~\}$ and
\eqn
w_\alpha(z_1,\cdots,z_N)&=&\prod_{m\in J^\alpha_-}~
B(z_1,\cdots,z_N |z_{\alpha_m})\Omega,\\
\Omega&=&v_+\otimes \cdots \otimes v_+
\quad \in V^{\otimes N}.
\endeqn
Then
\[
G(z_1,\cdots,z_N)=\sum_\alpha w_\alpha(z_1,\cdots,z_N)
H\bigl(\{z_m\}_{m\in J^{\alpha}_-}\,|\,\{z_p\}_{p\in
J^{\alpha}_+}\bigr)
\prod_{m\in J^{\alpha}_- \atop p\in J^{\alpha}_+}
{1 \over b(z_p/z_m)}
\,\prod_{m\in J^{\alpha}_-}\sqrt{z_m}.
\]
\bigskip

Under the relations (\ref{eqn:Rsym1})
and (\ref{eqn:Rsym}),
it is sufficient to consider the remaining relation
(\ref{eqn:cyclic})
for two cases with $\e_1=+$ and $\e_1=-$, e.g.,
$(\e_1,\cdots,\e_N)=
(+\underbrace{-\cdots-}_n
\underbrace{+\cdots  +}_{l-1})$ and
$(\underbrace{-\cdots-}_{n-1}\underbrace{+\cdots +}_l -)$.
Solving (\ref{eqn:Rsym}) for
the corresponding components
in terms of (\ref{eqn:1_comp}),
we find that the original system
(\ref{eqn:R-symm}--\ref{eqn:cyc})
is equivalent to the following
for the single function
$H(z_1,\cdots,z_n\,|\,z_{n+1},\cdots,z_N)$
satisfying the said symmetry condition:
\begin{equation}
\begin{array}{cl}
&\d_{+}^{-1} H
(z_2,\cdots,z_{n+1}|z_{n+2},\cdots,z_N,z_1\tau^4 ) \\
=&\displaystyle\prod_{j=2}^{n+1}
{z_1-z_j\tau^2 \over (z_1-z_j)\tau}
{}~H(z_2,\cdots,z_{n+1}|z_{n+2},\cdots,z_N,z_1) \\
-&\displaystyle\sum_{j=2}^{n+1}
{(1-\tau^2)z_1 \over (z_1-z_j)\tau}
\prod_{k=2 \atop k\neq j}^{n+1}
{z_j-z_k\tau^2 \over (z_j-z_k)\tau}
{}~H(z_1, \stackrel{j}{\hat{\cdots}},z_{n+1}|
z_j,z_{n+2},\cdots,z_N),
\end{array}
\label{eqn:B}
\end{equation}
\begin{equation}
\begin{array}{cl}
&\d_{-}\tau^{-2}\,
H(z_1\tau^{-4},z_2,\cdots,z_n|z_{n+1},\cdots,z_N) \\
=&
\displaystyle\prod_{j=n+1}^{N}
{z_1 -z_j \tau^{-2} \over (z_1 -z_j )\tau^{-1}}
{}~H(z_1,\cdots,z_n|z_{n+1},\cdots,z_N) \\
-&\displaystyle\sum_{j=n+1}^{N}
{(1-\tau^{-2})z_j \over (z_1 -z_j )\tau ^{-1}}
\prod_{k=n+1 \atop k\neq j}^{N}
{z_j -z_k \tau^{-2}\over (z_j -z_k )\tau ^{-1}}
{}~H(z_2,\cdots,z_n,z_j|z_1,z_{n+1},
\stackrel{j}{\hat{\cdots}},z_N).
\label{eqn:C}
\end{array}
\end{equation}

In what follows we tune $\d_+=\tau^{-n}$ and $\d_-=\tau^{-l}$.
We wish to find an integral formula of the form
\begin{equation}
H(z_1 , \cdots , z_N )=
(S_{n N} F)(z_1,\cdots ,z_N),
\end{equation}
where $S_{n N}$ stands for the integral transform
\begin{equation}
(S_{n N} F)(z_1,\cdots ,z_N)=
\prod_{\mu =1}^{n} \oint_{C} dx_{\mu }
F(x_1 , \cdots , x_n | z_1 , \cdots , z_N )
\Psi (x_1 , \cdots , x_n | z_1 , \cdots , z_N ).
\label{intform}
\end{equation}
The notation is explained below.

The kernel
$\Psi$
has the form
$$
\Psi (x_1 , \cdots , x_n | z_1 , \cdots , z_N )=
\vartheta (x_1 , \cdots , x_n | z_1 , \cdots , z_N )
\prod_{\mu =1}^{n}
\prod_{j=1}^{N} \psi \Bigl(\frac{x_{\mu }}{z_j }\Bigr),
$$
where
$$
\psi(z)=\frac{1}{(zq;q^4)_{\infty}
(z^{-1}q;q^4)_{\infty}},
\qquad
(z;p)_\infty=\prod_{n=0}^\infty(1-z p^n).
$$
For the function
$\vartheta$
we assume that
\begin{itemize}
\item
it is anti-symmetric and holomorhpic
in the $x_\mu \in \bc \backslash \{0\}$,

\item
it is symmetric and meromorphic
in the $\log z_j\in \bc$,

\item
it has  the transformation property
\begin{equation}
\begin{array}{rcl}
\vartheta (x_1 , \cdots , x_n |
z_1 , \cdots , z_j \tau^4 ,\cdots,z_N )
& = &
\vartheta (x_1 , \cdots , x_n | z_1 , \cdots , z_N )
\displaystyle\prod_{\mu =1}^{n}
\frac{-z_j \tau}{x_{\mu }}, \\
\vartheta (x_1 , \cdots, x_{\mu}\tau^4 ,
\cdots, x_n | z_1 , \cdots , z_N )
& = &
\vartheta (x_1 , \cdots , x_n | z_1 , \cdots , z_N )
\displaystyle\prod_{j=1}^{N} \frac{-x_{\mu } \tau}{z_j }.
\end{array}
\label{pstr}
\end{equation}
\end{itemize}
The function
$\vartheta$
is otherwise arbitrary,
and the choice of $\vartheta $'s corresponds to that of
solutions.
The transformation property of $\vartheta$ implies
\begin{eqnarray}
\frac{\Psi(x_1,\cdots,x_n|z_1,\cdots,z_j\tau^4,\cdots,z_N)}
{\Psi(x_1,\cdots,x_n|z_1,\cdots,z_N)}
&=&
\prod_{\mu=1}^n\frac{x_\mu-z_j\tau}{x_\mu-z_j\tau^3},
\label{eqn:trpsi1} \\
\frac{\Psi(x_1,\cdots,x_\mu\tau^4,\cdots,x_n|z_1,\cdots,z_N)}
{\Psi(x_1,\cdots,x_n|z_1,\cdots,z_N)}
&=&
\tau^{-2N}\prod_{j=1}^N\frac{x_\mu-z_j\tau^{-1}}{x_\mu-z_j\tau^{-3}}.
\label{eqn:trpsi2}
\end{eqnarray}
The integration $\oint_{C}dx_\mu$
is along a simple closed curve $C$ oriented
anti-clockwise,
which encircles the points
$z_j \tau ^{-1-4k} (1\leq j \leq N , k\geq 0)$
but not
$z_j \tau ^{1+4k} (1\leq j \leq N , k\geq 0)$.
Finally
\begin{equation}
F(x_1 , \cdots , x_n | z_1 , \cdots ,z_N )
=\frac{\Delta ^{(n l)}
(x_1 , \cdots , x_n | z_1 , \cdots , z_n |
                      z_{n+1}, \cdots , z_N )}
      {\displaystyle\prod_{j=1}^{n}\prod_{i=n+1}^{N}
       (z_i -z_j \tau ^2 )},
\label{Delta}
\end{equation}
where $\Delta^{(nl)}$ is
a certain homogeneous polynomial to be determined,
antisymmetric in the variables $(x_1 ,\cdots , x_n )$ and
symmetric in the variables $(z_1 , \cdots , z_n )$
and $(z_{n+1}, \cdots ,z_N)$ separately.

In the next sections we shall find the formula for it.

\section{The Case $n=1<l$}

In this section we find $\Delta=\Delta^{(1 l)}$ for $l>1$.
The result will be given in (\ref{eqn:n=1}).
First, we prepare some lemmas.
\begin{lem}~~~~Let
$f$ be a polynomial in $x$ and set
\begin{equation}
\begin{array}{l}
\displaystyle{\tilde f}(x|z_1,..,z_N)=
\displaystyle\frac{x-z_1 \tau }{x-z_1 \tau^3 }
f(x|z_1\tau^4,z_2,\cdots,z_N) \\
+\displaystyle \frac{z_1(-\tau)^{4-N}}{x\prod_{j=2}^N(z_j-z_1\tau^2)}
\left\{\frac{\prod_{j=1}^N(x-z_j \tau)}{x-z_1\tau^3}
-\tau^{2(N-2)}\prod_{j=2}^N(x-z_j\tau^{-1})\right\}
f(z_1\tau^3|z_1\tau^4,z_2,\cdots,z_N).\\
\end{array}\label{til}
\end{equation}
Then we have
\eq
(S_{1 N}f)(z_1\tau^4,z_2,\cdots,z_N)=
(S_{1 N}\tilde{f})(z_1,\cdots,z_N),
\label{eqn:S_n}
\endeq
where the LHS is the analytic continuation of
$\bigl(S_{1 N}f\bigr)(z_1,\cdots,z_N)$
in the variable $z_1$.
\end{lem}
{\sl Proof.}~~~~When the integral (\ref{eqn:S_n})
is analytically continued
from $z_1$ to $z_1 \tau^4$, the poles of the integrand move
from $x=\cdots,z_1\tau^{-5},z_1\tau^{-1},z_1\tau ,
z_1\tau^5,\cdots$ to
$x= \cdots,z_1\tau^{-1},z_1\tau^3,z_1\tau^5,z_1\tau^9,\cdots$.
In particular,
the pole which moves from $x=z_1\tau^{-1}$
to $x=z_1\tau^3$ crosses the original contour $C$.
Using (\ref{eqn:trpsi1}), we obtain
$$
(S_{1 N}f)(z_1\tau^4,z_2,\cdots,z_N)
=(\oint_C+2\pi i\hbox{\rm Res\/}_{x=z_1 \tau^3})
dx f(x|z_1\tau^4,z_2,\cdots,z_N)
\frac{x-z_1\tau}{x-z_1\tau^3} \Psi (x|z_1,\cdots,z_N).
$$

Set
$$
r(x)=\frac{z_1\tau^3}{x} \prod _{j=2}^N
\frac{x-z_j \tau}{z_1\tau^3-z_j\tau}.
$$
Then it has zeros at
$x=z_j\tau$ $(j=2,...,N)$ and
is equal to $1$ at $x=z_1\tau^3$
and therefore
the residue at $x=z_1 \tau ^3 $ can be replaced by
the difference of two integrals as follows:
$$
\begin{array}{cl}
&\displaystyle
2\pi i\hbox{\rm Res\/}_{x=z_1 \tau^3} dx
f(x|z_1\tau^4,z_2,\cdots,z_N)
\frac{x-z_1\tau}{x-z_1\tau^3}
\Psi(x|z_1,\cdots,z_N) \\
=&\left\{ \displaystyle\oint_{\tau^4C} -\oint_C \right\}
dx f(z_1\tau^3|z_1\tau^4,z_2,\cdots,z_N)
{\displaystyle \frac{x-z_1\tau}{x-z_1\tau^3} }
r(x) \Psi (x|z_1\cdots,z_N)\\
=&-\displaystyle\oint_C dx
\left\{
\frac{x-z_1\tau}{x-z_1\tau^3}r(x)-
\tau^{-2(N-2)}
\frac{x\tau^4-z_1\tau}{x\tau^4-z_1\tau^3}
\prod_{j=1}^N\frac{ x-z_j\tau^{-1}}{x-z_j\tau^{-3}}
r(x\tau^4)\right\}\times\\
\times & f(z_1 \tau^3|z_1 \tau^4,z_2,\cdots,z_N)
\Psi(x|z_1,\cdots,z_N).
\end{array}
$$
In the second equality we use (\ref{eqn:trpsi2}).
Thus we obtain (\ref{eqn:S_n}).~~~~$\Box$

{}~

\begin{lem}~~~~If
$f(x|z_1,\cdots,z_N)$ is a polynomial in
$x$, then there exists $g(x|z_1,\cdots,z_N)$,
a polynomial in $x$ of degree less than or equal to $N-2$, such that
$S_{1N}f=S_{1N}g$.
\label{hN}
\end{lem}
{\sl Proof.}~~~~
Since the product $\prod_{j=1}^N(x-z_j \tau )$
cancels the simple poles at $x=z_j \tau $ of
$\Psi (x|z_1,\cdots,z_N)$, we can deform
the contour $C$ to $\tau^4C$ without crossing
the poles as follows:
$$
\begin{array}{rcl}
\displaystyle\oint_C dx
x^{k-1} \prod_{j=1}^N(x-z_j \tau )
\Psi (x|z_1,\cdots,z_N)
&=&\displaystyle\oint_{\tau^4 C} dx
x^{k-1} \prod_{j=1}^N (x-z_j \tau ) \Psi(x|z_1,\cdots,z_N) \\
&=&\displaystyle\tau^{2N+4k}
\oint_C dx x^{k-1}\prod_{j=1}^N(x-z_j\tau^{-1} )
\Psi(x|z_1,\cdots,z_N).
\end{array}
$$
This implies that
$$
\oint_C dx
f(x) \Psi (x|z_1 , \cdots , z_N )=0
$$
for the polynomials
$$
f(x)=x^{k-1}
\left\{\prod_{j=1}^N(x-z_j\tau )-\tau^{2N+4k}
\prod_{j=1}^{N}(x-z_j\tau^{-1})\right\}, \quad k\geq 0.
$$
{}From these relations any polynomial is reducible to
some $g$ of degree less than $N-1$.~~~~$\Box$.

{}~
We are now to solve (\ref{eqn:B})
in the form $H=S_{1N}F$, where $F$ is given
by (\ref{Delta}) with $n=1$.
Because of (\ref{eqn:S_n}),
the LHS of (\ref{eqn:B}) is equal to $\tau S_{1N}{\tilde{F_1}}$
where $F_1(x|z_1,\cdots,z_N)=F(x|z_2,\cdots,z_N,z_1)$
and ${\tilde{F_1}}$ is
given by (\ref{til}) with $f$ replaced by $F_1$.
Similarly, the RHS of (\ref{eqn:B})
is equal to $S_{1N}F_2$ for a certain $F_2$.
Because of Lemma \ref{hN},
we make an Ansatz that the equation
$\tau{\tilde{F_1}}=F_2$ is satisfied by an $F$ such that
$\Delta$ is a homogeneous polynomial of
degree $N-2$.
In fact,
this Ansatz uniquely determines
$\Delta$ as we will see shortly.

{}~

Define
\begin{equation}
h^{(N)} (x|z_1 , \cdots , z_N ):=\frac{1}{x}
\left\{ \prod_{j=1}^{N} (x-z_j \tau ) -
\tau ^{2N} \prod_{j=1}^{N} (x-z_j \tau ^{-1}) \right\}.
\label{defH}
\end{equation}
Eq. (\ref{eqn:B}) for $\Delta$ reads
$$
\begin{array}{cl}
\displaystyle
&\displaystyle\frac{1}
{(z_2-z_1\tau^2)\prod_{j=3}^{N}(z_j-z_2\tau^2)}
\frac{x-z_1\tau}{x-z_1\tau^3}
\Delta(x|z_2|z_3,\cdots,z_N,z_1\tau^4)
\\
+&\displaystyle\frac{
(-\tau)^{4-N}z_1
\Delta(z_1\tau^3|z_2|z_3,\cdots,z_N,z_1\tau^4)}
{x(z_2-z_1\tau^2)^2
\prod_{j=3}^{N}(z_j-z_1\tau^2)
\prod_{j=3}^N(z_j-z_2\tau^2)}
\left\{\frac{\prod_{j=1}^{N}(x-z_j\tau)}{x-z_1\tau^3}
-\tau^{2(N-2)}\prod_{j=2}^{N}(x-z_j\tau^{-1})\right\}
\\
=&\displaystyle\frac{1}{z_2-z_1}
\left\{\frac{1}{\prod_{j=3}^N(z_j-z_2\tau^2)}
\Delta(x|z_2|z_3,\cdots,z_N,z_1)
-\frac{(1-\tau^2)z_1}{\prod_{j=2}^N (z_j-z_1\tau^2 )}
\Delta(x|z_1|z_2,\cdots,z_N)\right\}.
\end{array}
$$
Comparing the residues at $z_3=z_1\tau^2$, we obtain
$$
\begin{array}{rcl}
\displaystyle \Delta
(x|z_1|z_2,z_1\tau^2,z_4,...,z_N)&=&
\displaystyle\frac{(-\tau)^{2-N}(x-z_1\tau)}{(1-\tau^2)(z_2-z_1\tau^2)}
\prod_{j=4}^N \frac{1}{z_j-z_2\tau^2}
h^{(N-2)}(x|z_2,z_4,...,z_N)\times\\
& \times & \displaystyle\Delta
(z_1\tau^3|z_2|z_1\tau^2,z_4,...,z_N,z_1\tau^4).
\end{array}
$$
This determines the restriction
of $\Delta$ at $z_3=z_1\tau^2$
up to a constant multiple.
Choosing the constant appropriately we have
$$
\Delta
(x|z_1|z_2,z_1\tau^2,z_4,\cdots,z_N)=
(x-z_1 \tau )h^{(N-2)}(x|z_2,z_4,...,z_N).
$$
Because of the symmetry of $\Delta(x|z_1|z_2,\cdots,z_N)$ with
respect to $(z_2,\cdots,z_N)$, we have a similar equation for
the restriction at $z_j=z_1\tau^2$ for $2\leq j\leq N$.
These $N-1$ restrictions uniquely determine the degree $N-2$ polynomial
$\Delta (x|z_1|z_2,\cdots,z_N)$:
\begin{equation}
\Delta (x|z_1|z_2,\cdots,z_N)
=(x-z_1\tau)\sum_{\kappa=0}^{N-3}(1-\tau^{2(N-2-\kappa)})
(-\tau)^\kappa x^{N-3-\kappa}\sum_{\lambda=0}^\kappa(-z_1\tau^2)^\lambda
\sigma_{\kappa-\lambda}(z_2 \cdots z_N),
\label{eqn:n=1}
\end{equation}
where $\sigma_\kappa(a_1,\cdots,a_n)$ denotes the $\kappa$-th
elementary symmetric polynomial:
\[
\prod_{j=1}^{n} (t-a_j )  =
\sum_{\kappa =0}^{n} (-1)^{\kappa }
\sigma _{\kappa } (a_1 , \cdots ,a_n ) t^{n-\kappa }.
\]

To prove that this $\Delta$ actually solves
(\ref{eqn:B}) and (\ref{eqn:C}), we use the following simple
facts:
Let $P,Q$ be homogeneous rational functions of multi-variables.
Suppose that the poles of $P,Q$ are simple and
they are contained in
a union of $k$ hyperplanes $H_j\hskip5pt(1\leq j\leq k)$.
Then, we can conclude $P=Q$ in each of the following:
\begin{itemize}
\item
The degrees of $P,Q$ are less than or equal to $-1$ and all the residues
of $P,Q$ at $H_j$'s coincide.
\item
The degrees of $P,Q$ are less than or equal to $-m$ and
the residues of $P,Q$ coincide at $H_j$ for at least $(k-m+1)$ of them.
\end{itemize}

{}~
\section{The General Case }

We now present
our integral formula for the case $n<l$.
The case $n=l$ will be discussed at the end of this section.
The polynomial $\Delta ^{(n l)}$ in (\ref{Delta})
is given by
\begin{equation}
\Delta ^{(n l)}(x_1 , \cdots , x_n | z_1 , \cdots , z_n |
                      z_{n+1}, \cdots , z_N )
=\det \left( A_{\lambda }^{(n l)} (x_{\mu }|
z_1 , \cdots , z_n |
                      z_{n+1}, \cdots , z_N )
\right) _{1\leq \lambda , \mu \leq n}.
\label{determ}
\end{equation}
The entries of the $n\times n$ matrix $A^{(n l)}$
is defined as follows.
Let us introduce the  polynomials
$\tilde{\sigma }_{\kappa } (a_1 , \cdots , a_n )$  by
$$
\prod_{j=1}^{n} (t-a_j )^{-1}  =
\sum_{\kappa \geq 0} \tilde{\sigma }_{\kappa }
(a_1 , \cdots , a_n )
t^{-n-\kappa }.
$$
For $\lambda ,n, l \geq 0$,
define the following polynomials:
$$
f_{\lambda }^{(n l)} (x| a_1 , \cdots , a_n |
b_1 , \cdots , b_l )=
\sum_{\kappa =0}^{l-n+\lambda -2} (1-\tau ^{2(l-n+\lambda -1-\kappa )})
(-\tau )^{\kappa }
\varphi _{\lambda \kappa }^{(n l)}(a_1 , \cdots , a_n | b_1, \cdots , b_l )
x^{l-n+\lambda -2-\kappa },
$$
where
$$
\begin{array}{l}
\varphi _{\lambda \kappa }^{(n l)}(a_1 , \cdots , a_n | b_1, \cdots , b_l )
= \sigma _{\kappa } (b_1 , \cdots , b_l ) \\
- \displaystyle\sum _{\lambda \leq \alpha \leq \beta \leq \kappa }
(-1)^{\beta -\alpha }
\tau ^{2\beta }
\sigma _{\kappa -\beta } (b_1 , \cdots , b_l )
\sigma _{\alpha }(a_1 , \cdots , a_n )
\tilde{\sigma }_{\beta -\alpha } (a_1 , \cdots , a_n ),
\end{array}
$$
and
$$
g_{\lambda }^{(n)} (x|a_1 , \cdots , a_n )=
\sum_{\kappa =0}^{\lambda -2} (1-\tau ^{2(\lambda -1-\kappa )})
(-\tau )^{\kappa } \sigma _{\kappa } (a_1 , \cdots , a_n )
x^{\lambda -2-\kappa }.
$$
Note that $\varphi ^{(n l)}_{\lambda \kappa }
(a_1 ,\cdots , a_n | b_1 , \cdots , b_l )=
\sigma _{\kappa }(b_1 , \cdots , b_l )$
if $\kappa < \lambda $ or $\lambda >n$.
For $\lambda \geq 0$ and $n<l$, set
\begin{equation}
\begin{array}{rl}
A_{\lambda }^{(n l)} (x| a_1 , \cdots , a_n |
b_1 , \cdots b_l ):= &
\displaystyle\prod_{j=1}^{n} (x-a_j \tau )
f_{\lambda }^{(n l)} (x| a_1 , \cdots , a_n | b_1 , \cdots ,b_l ) \\
& + \tau ^{2(l-n+\lambda -1)}
\displaystyle\prod_{i=1}^{l} (x-b_i \tau ^{-1})
g_{\lambda }^{(n)} (x| a_1 , \cdots , a_n ).
\label{defA}
\end{array}
\label{nnA}
\end{equation}
This is a homogeneous polynomial of degree $l+\lambda -2$,
symmetric with respect to $a_j $'s and $b_i $'s separately.
Notice that
\begin{equation}
A_{\lambda }^{(n l)} (x| a_1 , \cdots , a_n |
b_1 , \cdots b_l )
\mbox{ is linear with respect to $b_i $'s.}
\label{blin}
\end{equation}
By the construction (\ref{determ}) and (\ref{nnA})
$\Delta ^{(n l)}$
is a homogeneous polynomial of degree $n(l-1)+n(n-1)/2$
with correct symmetries.

{}~

\begin{lem}~~~~The polynomial
$A_{\lambda }^{(n l)}$ obeys the following
recursion relation
$$
\begin{array}{cl}
& A_{\lambda }^{(n l)} (x| a_1 , \cdots , a_{n-1}, a |
b_1 , \cdots ,b_{l-1}, a\tau ^2 )
= (x-a\tau ) ~\times \\
\times &
\left\{ A_{\lambda }^{(n-1 l-1)}
(x| a_1 , \cdots , a_{n-1} | b_1 , \cdots ,b_{l-1})-
a\tau ^3 A_{\lambda -1}^{(n-1 l-1)}
(x| a_1 , \cdots , a_{n-1} | b_1 , \cdots ,b_{l-1}) \right\}.
\end{array}
$$
\label{recA}
\end{lem}
{\sl Proof.}~~~~
It follows from the recursions for $f_{\lambda }^{(n l)}$ and
$g_{\lambda }^{(n)}$
$$
\begin{array}{rl}
f_{\lambda }^{(n l)} (x| a_1 , \cdots , a_{n-1}, a |
b_1 , \cdots ,b_{l-1}, a\tau ^2 )=&
f_{\lambda }^{(n-1 l-1)}
(x| a_1 , \cdots , a_{n-1} | b_1 , \cdots ,b_{l-1}) \\
-&
a\tau ^3 f_{\lambda -1}^{(n-1 l-1)}
(x| a_1 , \cdots , a_{n-1} | b_1 , \cdots ,b_{l-1}), \\
g_{\lambda }^{(n)} (x| a_1 , \cdots , a_{n-1}, a)=&
g_{\lambda }^{(n-1)} (x| a_1 , \cdots , a_{n-1})-
a\tau g_{\lambda -1}^{(n-1)} (x| a_1 , \cdots , a_{n-1}).~~~~\Box
\end{array}
$$

{}~

\begin{lem}~~~~The determinant
$\Delta^{(n l)}$ obeys the following
recursion relation
\begin{equation}
\begin{array}{cl}
& \Delta^{(n l)} (x_1 , \cdots , x_n
| a_1 , \cdots , a_{n-1}, a |
b_1 , \cdots ,b_{l-1}, a\tau ^2 )
= \displaystyle\prod_{\mu =1}^n (x_{\mu }-a\tau )
\sum_{\nu =1}^n (-1)^{n+\nu } ~\times \\
\times & \displaystyle h^{(N-2)} (x_{\nu }|
a_1 , \cdots , a_{n-1}, b_1 , \cdots ,b_{l-1})
\Delta^{(n-1 l-1)} (x_1 , \stackrel{\nu }{\hat{\cdots }}, x_n
| a_1 , \cdots , a_{n-1}|
b_1 , \cdots ,b_{l-1}),
\end{array}
\label{recD}
\end{equation}
where $h^{(N)}$ is defined in (\ref{defH}).
\label{Del}
\end{lem}
{\sl Proof.}~~~~
By the definition we get
$$
A_{0}^{(n l)} (x| z_1 , \cdots , z_{n} |
z_{n+1} , \cdots , z_{N})=0,
$$
and
$$
A_{n+1}^{(n l)} (x| z_1 , \cdots , z_{n} |
z_{n+1} , \cdots ,z_{N})=h^{(N)}(x|z_1 , \cdots , z_N ).
$$
Using Lemma \ref{recA} we have
$$
\Delta^{(n l)} (x_1 , \cdots , x_n
| a', a |
b', a\tau ^2 )
=\displaystyle\prod_{\mu =1}^n (x_{\mu }-a\tau )~ {\rm det}
\left(
A_{\lambda }^{(n-1 l-1)} (x_{\mu }|a'|b')-a\tau ^3
A_{\lambda -1}^{(n-1 l-1)} (x_{\mu }|a'|b') \right)_{
1 \leq \lambda , \mu \leq n},
$$
where we employ the
abbreviations $a'=(a_1 , \cdots , a_{n-1})$ and
$b'=(b_1 , \cdots , b_{l-1})$.
Multiply the $\lambda $-th row
by $a\tau ^3 $ and add it to the $(\lambda +1)$-th row
successively for $\lambda =1, \cdots ,n-1$.
Then we obtain (\ref{recD}).~~~~$\Box$

{}~

{}In terms of $\Delta^{(nl)}$ we have
\begin{equation}
\begin{array}{rl}
& {\rm RHS~ of ~(\ref{eqn:B})}
=\displaystyle\prod_{j=2}^{n+1} (z_j -z_1 \tau ^2 )^{-1}
\displaystyle\prod_{j=2}^{n+1}\displaystyle\prod_{i=n+2}^{N}
(z_i -z_j \tau ^2 )^{-1} \times \\
\times & \displaystyle\prod_{\mu =1}^{n}
\displaystyle\oint_{C} dx_{\mu }
\hat{\Delta }^{(n l)}(x_1 , \cdots , x_n |
                        z_1 | z_2 , \cdots , z_{n+1} |
                        z_{n+2}, \cdots , z_N )
\Psi (x_1 , \cdots , x_n | z_1 , \cdots , z_N ),
\end{array}
\end{equation}
where
\begin{eqnarray}
&& \hat{\Delta }^{(n l)}(x_1 , \cdots , x_n |
                        z_1 | z_2 , \cdots , z_{n+1} |
                        z_{n+2}, \cdots , z_N ) \nonumber \\
& = & \displaystyle\prod_{j=2}^{n+1}
 \frac{z_j -z_1 \tau ^2 }{(z_1 -z_j )\tau }
\Delta ^{(n l)}
(x_1 , \cdots , x_n | z_2 , \cdots , z_{n+1} |
                      z_{n+2}, \cdots , z_N , z_1 )
\label{tiDel} \\
& - & \displaystyle\sum_{j=2}^{n+1}
 \frac{(1-\tau ^2 )z_1 }{(z_1 -z_j )\tau }
 \prod_{k=2 \atop k\neq j}^{n+1}
 \frac{z_k -z_1 \tau ^2 }{(z_j -z_k )\tau }
 \prod_{i=n+2}^{N}
 \frac{z_i -z_j \tau ^2 }{z_i -z_1 \tau ^2 }
\Delta ^{(n l)}
(x_1 , \cdots , x_n |
z_1 , z_2 , \stackrel{j}{\hat{\cdots }}, z_{n+1} |
                      z_{n+2}, \cdots , z_N , z_j ). \nonumber
\end{eqnarray}

By repeating the argument of Lemma \ref{eqn:S_n}
we find for $H=H^{(nl)}$
\begin{equation}
\begin{array}{cl}
&H^{(n l)}
(z_2 , \cdots ,z_{n+1} |
                      z_{n+2}, \cdots , z_{N}, z_1 \tau ^4 ) \\
=&
\displaystyle\prod_{j=2}^{n+1} (z_1 \tau ^4 -z_j \tau ^2 )^{-1}
\displaystyle\prod_{j=2}^{n+1}\displaystyle\prod_{i=n+2}^{N}
(z_i -z_j \tau ^2 )^{-1}
I^{(n l)}
(z_2 , \cdots ,z_{n+1} |
                      z_{n+2}, \cdots , z_{N}, z_1 \tau ^4 ),
\end{array}
\end{equation}
where
\begin{eqnarray}
&&I^{(n l)}
(z_2 , \cdots ,z_{n+1} |
z_{n+2}, \cdots , z_{N}, z\tau ^4 )\\
&=&
\displaystyle\prod_{\mu =1}^{n}
\displaystyle\oint_{C} dx_{\mu }
\mbox{det} \left(
\widetilde{A}_{\lambda }^{(n l)}
(x_{\mu }| z_2 , \cdots , z_{n+1} |
z_{n+2} , \cdots, z_N , z_1 \tau ^4 )
\right) _{1\leq \lambda , \mu \leq n}
\Psi (x_1 , \cdots , x_n | z_1 , \cdots , z_N ), \nonumber
\end{eqnarray}
and
\begin{equation}
\begin{array}{cl}
&\displaystyle
\widetilde{A}_{\lambda }^{(n l)}
(x|z_2 , \cdots , z_{n+1}
| z_{n+2}, \cdots , z_N ,
z_1 \tau ^4) \\
=&
\displaystyle\frac{x-z_1 \tau }{x-z_1 \tau ^3}
A_{\lambda }^{(n l)} (x|
z_2 , \cdots , z_{n+1}|
z_{n+2}, \cdots , z_N , z_1 \tau ^4) \\
+&
\displaystyle\frac{z_1 (-\tau )^{4-N}}{x}
\left\{
\frac{x-z_1 \tau }{x-z_1 \tau ^3}
\prod_{j=2}^{N} \frac{x-z_j \tau }
{z_j -z_1 \tau ^2 }
-\tau ^{2(N-2)}
\prod_{j=2}^{N}
\frac{x-z_j \tau ^{-1}}{z_j -z_1 \tau ^2 }
\right\} \times \\
\times & A_{\lambda }^{(n l)} (z_1 \tau ^3 |
z_2 , \cdots , z_{n+1}|z_{n+2}, \cdots , z_N ,
 z_1 \tau ^4 ).
\end{array}
\label{DA}
\end{equation}

Let us prove (\ref{eqn:B}) and (\ref{eqn:C}) for $H^{(nl)}$.
The first equation (\ref{eqn:B})
is satisfied if the following proposition holds:

\begin{prop}~~~~Let
$\widetilde{A}_{\lambda }^{(n l)}$ and
$\hat{\Delta }^{(n l)}$ be defined by
(\ref{DA}) and (\ref{tiDel}), respectively.
Then they satisfy the following equation
\begin{equation}
\begin{array}{rl}
&{\rm det} \left(
\widetilde{A}_{\lambda }^{(n l)} (x_{\mu }| z_2 , \cdots , z_{n+1} |
z_{n+2} , \cdots, z_N , z_1 \tau ^4 )
\right) _{1\leq \lambda , \mu \leq n} \\
=&
(-\tau )^n
\hat{\Delta }^{(n l)}(x_1 , \cdots , x_n |
                        z_1 | z_2 , \cdots , z_{n+1} |
                        z_{n+2}, \cdots , z_N ).
\end{array}
\label{DAB}
\end{equation}
\label{main}
\end{prop}
{\sl Proof.}~~~~In this proof
we use the abbreviations
$x=(x_{1}, \cdots , x_n )$,
$\stackrel{\nu }{\hat{x}}=
(x_1 , \stackrel{\nu }{\hat{\cdots }}, x_n )$,
$z'=(z_2 , \cdots , z_{n+1})$,
$\stackrel{j}{\hat{z'}}=
(z_2 , \stackrel{j}{\hat{\cdots }}, z_{n+1})$,
$z''=(z_{n+2}, \cdots , z_N )$
and
$\stackrel{i}{\hat{z''}}=
(z_{n+2}, \stackrel{i}{\hat{\cdots }}, z_{N})$.
First of all, observe
$$
\widetilde{A}_{\lambda }^{(n l)} (x| z'| z'', z_1 \tau ^4 ) =
\frac{x-z_1 \tau }{x-z_1 \tau ^3 }
A_{\lambda }^{(n l)} (x| z' |  z'', z_1 \tau ^4 )
+g(x|z_1 |z'|z'')f_{\lambda }^{(n l)} (z_1 \tau ^3 |
z'|z'', z_1 \tau ^4 ),
$$
where
$$
g(x|z_1 |z'|z''):=\frac{z_1 (-\tau )^{4-l}}{x}
\left\{
\frac{x-z_1 \tau }{x-z_1 \tau ^3 }
\prod_{j=2}^{N} (x-z_j \tau )-\tau ^{2(N-2)}
\prod_{j=2}^{N} (x-z_j \tau ^{-1}) \right\}
\left/ \prod_{i=n+2}^{N}(z_i -z_1 \tau ^2 ).\right.
$$
Thanks to the $n$-fold linearity of the determinant
we have
\begin{equation}
\begin{array}{cl}
&\det \left(
\widetilde{A}_{\lambda }^{(n l)}
(x_{\mu }| z_2 , \cdots , z_{n+1} |
z_{n+2} , \cdots, z_N , z_1 \tau ^4 )
\right) _{1\leq \lambda , \mu \leq n} \\
=&\displaystyle\prod_{\mu =1}^{n}
\frac{x_{\mu }-z_1 \tau }{x_{\mu }-z_1 \tau ^3 }
{\rm det}\left(
A_{\lambda }^{(n l)} (x_{\mu }| z'|
z'', z_1 \tau ^4 ) \right) _{
1 \leq \lambda , \mu \leq n} \\
+&\displaystyle\prod_{\mu =1}^n (x_{\mu }-z_1 \tau )
\sum_{\nu =1 }^n
\prod_{\mu =1 \atop \mu \neq \nu }^n
\frac{g(x_{\nu }|z_1 |z'|z'')}{x_{\nu }-z_1 \tau ^3 }
{\rm det}\left(
B_{\lambda \mu }^{(n l)} \right) _{
1 \leq \lambda , \mu \leq n},
\label{n-fold}
\end{array}
\end{equation}
where
\begin{equation}
B_{\lambda \mu }^{(n l)}=\left\{
\begin{array}{ll}
\displaystyle A_{\lambda }^{(n l)}
{}~(x_{\mu }|z'|z'', z_1 \tau ^4 )/
(x_{\mu }-z_1 \tau ^3 ), &
\mbox{if $\lambda \neq \nu $,}\\
f_{\lambda }^{(n l)} (z_1 \tau ^3 |
z'|z'', z_1 \tau ^4 ),
& \mbox{if $\lambda =\nu $.}
\end{array}
\right.
\label{eqn:B=}
\end{equation}

Both sides of (\ref{DAB}) are
anti-symmetric polynomials
in the variables $(x_1 , \cdots , x_n )$
and symmetric rational functions in the variables
$(z_2 , \cdots , z_{n+1})$ and $(z_{n+2}, \cdots , z_N )$,
respectively.
The homogeneous degree of them is $n(l-1)+n(n-1)/2$.
{}From (\ref{tiDel}) and (\ref{n-fold}--\ref{eqn:B=})
all the singularities come from simple poles
located at
$z_i =z_1 \tau ^2$ ($n+1 \leq i \leq N$).

Let $Cyc (n,l)$ denote the statement that
eq. (\ref{DAB}) holds for $(n,l)$ satisfying $n<l$.
Then in order to verify $Cyc (n,l)$
it is enough to show the following three claims:

Claim 1.~~~~When $z_j =z_1 \tau ^2$ ($2\leq j \leq n+1$),
 (\ref{DAB}) holds.

Claim 2.~~~~When $z_i =z_j \tau ^2$ ($2\leq j \leq n+1 < i \leq N$),
 (\ref{DAB}) holds.

Claim 3.~~~~The residues at $z_i =z_1 \tau ^2$  ($n+2 \leq i \leq N$)
in both sides of (\ref{DAB}) are equal.

\noindent This sufficiency is based on
an elementary algebra.
If these three are valid,
the difference of both sides is a polynomial
of degree $n(l-1)+n(n-1)/2$
because of Claim 3;
it has $nl+n(n-1)/2$ zeros
because of
the first two claims and the anti-symmetry
with respect to $x_{\mu }$'s,
and consequently it should vanish.

Let us show $Cyc (n,l)$ by induction.

We have proved $Cyc (1,l)$
in section 3.

For $n<l$ consider (\ref{DAB})
under the restriction of $z_j =z_1 \tau ^2$ ($2\leq j \leq n+1$).
Then we get
$$
{\rm RHS}|_{z_j =z_1 \tau ^2 }
=
\prod_{i=n+2}^N \frac{z_i -z_1 \tau ^4 }{z_i -z_1 \tau ^2 }
{}~
\Delta ^{(n l)}
(x|
z_1 , \stackrel{j}{\hat{z'}} |
z'' , z_j )|_{z_j =z_1 \tau ^2 }.
$$
In the LHS,
we use the recursion relations of
$A_{\lambda }^{(n l)}$
and
$f_{\lambda }^{(n l)}$ and subtract
the terms
$A_{\lambda -1}^{(n-1 l-1)}$
and
$f_{\lambda -1}^{(n-1 l-1)}$ appearing in
the $\lambda $-th row, to find
$$
{\rm LHS}|_{z_j =z_1 \tau ^2 } =\prod_{\mu =1}^n (x_{\mu } -z_1 \tau )
{}~\mbox{det}\left(
A_{\lambda }^{(n-1 l-1)} (x_{\mu }|\stackrel{j}{\hat{z'}}|z'')+
\tilde{g}(x_{\mu })
f_{\lambda }^{(n-1 l-1)} (z_1 \tau ^3 |\stackrel{j}{\hat{z'}}|z'')
\right)_{
1 \leq \lambda , \mu \leq n},
$$
where
$$
\tilde{g}(x)=
z_1 (-\tau )^{4-l} h^{(N-2)}
(x|\stackrel{j}{\hat{z'}},z'')
\left/ \prod_{i=n+2}^N (z_i -z_1 \tau ^2 ). \right.
$$
Since the $n$-th row is equal to $h^{(N-2)}
(x_{\mu }|\stackrel{j}{\hat{z'}},z'') \displaystyle\prod_{
i=n+2}^N (z_i -z_1 \tau ^4 )/(z_i -z_1 \tau ^2 )$,
the second term \\
$\tilde{g}(x_{\mu })
f_{\lambda }^{(n l)} (z_1 \tau ^3 |\stackrel{j}{\hat{z'}}|z'')$
in the
$\lambda$-th row can be removed for $\lambda\neq n$.
Hence
we obtain
$$
{\rm LHS}|_{z_j =z_1 \tau ^2 }
=\displaystyle\prod_{\mu =1}^n (x_{\mu } -z_1 \tau )
\prod_{
i=n+2}^N \frac{z_i -z_1 \tau ^4 }{z_i -z_1 \tau ^2 }
\sum_{\nu =1}^n (-1)^{n+\nu }
\displaystyle h^{(N-2)} (x_{\nu }|
\stackrel{j}{\hat{z'}}|z'')
\displaystyle\Delta^{(n-1 l-1)}
(\stackrel{\nu }{\hat{x}}
|\stackrel{j}{z'}| z'').
$$
Thus Claim 1 is proved.

Next let us show Claim 2.
Suppose $Cyc(n-1,l-1)$.
The value of the RHS
at $z_i =z_j \tau^2 (2\leq j \leq n+1<i\leq N)$
is given recursively by
$$
\begin{array}{cl}
&{\rm RHS}|_{
z_i =z_j\tau ^2 }=(-\tau )^{n-1}
\displaystyle\frac{z_j -z_1 \tau ^2 }{z_j -z_1 }
\displaystyle\prod_{\mu =1}^n (x_{\mu }-z_j \tau )
\times \\
\times &
\displaystyle\sum_{\nu =1}^n (-1)^{n+\nu }
h^{(N-2)} (x_{\nu }| z_1 ,
\stackrel{j}{\hat{z'}}, \stackrel{i}{\hat{z''}})
\tilde{\Delta }^{(n-1 l-1)}
(\stackrel{\nu }{\hat{x}}, |
z_1 | \stackrel{j}{\hat{z'}}|\stackrel{i}{\hat{z''}}).
\end{array}
$$
Repeating a similar calculation
as before we have
$$
\begin{array}{cl}
&{\rm LHS}|_{z_i =z_j \tau ^2 }=
\displaystyle\frac{z_j -z_1 \tau ^2 }{z_j -z_1}
\prod_{\mu =1}^n (x_{\mu }-z_j \tau ) ~\times \\
\times & \displaystyle\sum_{\nu =1}^n
(-1)^{n+\nu }
h^{(N-2)} (x_{\nu }| z_1 ,
\stackrel{j}{\hat{z'}}, \stackrel{i}{\hat{z''}})~
{\rm det} \left(
\widetilde{A}_{\lambda }^{(n-1 l-1)}
(x_{\mu }|
\stackrel{j}{\hat{z'}}| \stackrel{i}{\hat{z''}}, z_1 \tau ^4 )
\right) _{1\leq \lambda \leq n-1
\atop 1\leq \mu (\neq \nu ) \leq n}.
\end{array}
$$
Thus Claim 2 follows from
the assumption of induction.

Let us turn to Claim 3.
It follows from (\ref{tiDel})
that the residue of
the RHS at $z_i =z_1 \tau ^2 $ is given by
$$
\begin{array}{cl}
&(-1)^{n+1}z_1 \tau ^2(1-\tau ^2 )
\displaystyle\prod_{\mu =1}^n (x_{\mu }-z_1 \tau )
\displaystyle\sum_{j=2}^{n+1}
 \prod_{k=2 \atop k\neq j}^{n+1}
 \frac{z_k -z_1 \tau ^2 }{(z_j -z_k )\tau }
 \prod_{m=n+2 \atop m \neq i}^{N}
 \frac{z_{m} -z_j \tau ^2 }{z_{m} -z_1 \tau ^2 } \times \\
\times &
\displaystyle\sum_{\nu =1 }^n (-1)^{n+\nu }
h^{(N-2)} (x_{\nu }|
z', \stackrel{i}{\hat{z''}})~
\Delta ^{(n-1 l-1)}
(\stackrel{\nu }{\hat{x}}|
\stackrel{j}{\hat{z'}}|\stackrel{i}{\hat{z''}}, z_j ).
\end{array}
$$
{}From (\ref{n-fold}) that of the LHS is equal to
$$
\displaystyle z_1 (-\tau )^{4-l}
\prod_{\mu =1}^{n} (x_{\mu }-z_1 \tau )
\displaystyle\sum_{\nu =1 }^n
h^{(N-2)} (x_{\nu }|
z', \stackrel{i}{\hat{z''}})~
{\rm det}\left( B'^{(n l)}_{\lambda \mu }
\right) _{1 \leq \lambda , \mu \leq n}\left/
\displaystyle\prod_{m=n+2 \atop m \neq i}^N
(z_{m}-z_1 \tau ^2 ), \right.
$$
where $B'^{(n l)}$ is obtained
from $B^{(n l)}$ by the
elementary transformation
\begin{equation}
B'^{(n l)}_{\lambda \mu }=
\left\{ \begin{array}{ll}
\displaystyle\prod_{j=2}^{n+1} (x_{\mu }-z_j \tau )
\frac{f_{\lambda }^{(n l)} (x_{\mu }
|z'|z'',z_1 \tau ^4 )|_{z_i =z_1 \tau ^2 }
-
f_{\lambda }^{(n l)} (z_1\tau^3
|z'|z'',z_1 \tau ^4 )|_{z_i =z_1 \tau ^2 }}
{x_{\mu }-z_1 \tau }  & \\
+\tau ^{2(l-n+\lambda -1)} (x_{\mu }-z_1 \tau )
\displaystyle\prod_{m=n+2 \atop m \neq i}^{N}
(x_{\mu }-z_{m} \tau ^{-1})
g_{\lambda }^{(n)} (x_{\mu }|z' ), &
\mbox{if $\mu \neq \nu $,} \\
f_{\lambda }^{(n l)}
(z_1 \tau ^3 |z'|z'',z_1 \tau ^4 )|_{z_i =z_1 \tau ^2 }, &
\mbox{if $\mu =\nu $.}
\end{array}
\right.
\label{Atr}
\end{equation}
Hence Claim 3 is equivalent to
\begin{equation}
\begin{array}{cl}
&
\displaystyle\sum_{\nu =1 }^n
h^{(N-2)} (x_{\nu }|
z', \stackrel{i}{\hat{z''}})~
{\rm det}\left(
B'^{(n l)}_{\lambda \mu }
\right) \\
=&(-1)^{N-1} \tau ^{l-2} (1-\tau ^2 )
\displaystyle\sum_{j=2}^{n+1}
 \prod_{k=2 \atop k\neq j}^{n+1}
 \frac{z_k -z_1 \tau ^2 }{(z_j -z_k )\tau }
 \prod_{m=n+2 \atop m \neq i}^{N}
 (z_{m} -z_j \tau ^2)  \times \\
\times&\displaystyle\sum_{\nu =1 }^n (-1)^{n+\nu }
h^{(N-2)} (x_{\nu }|
z', \stackrel{i}{\hat{z''}})~
\Delta ^{(n-1 l-1)}
(\stackrel{\nu }{\hat{x}}|
\stackrel{j}{\hat{z'}}|\stackrel{i}{\hat{z''}}, z_j ).
\end{array}
\label{res}
\end{equation}
The poles at $x_{\mu }=z_1 \tau $ ($1\leq \mu \leq n$)
in the LHS
and at $z_j =z_k$ ($2\leq j,k \leq n+1$) in RHS
are spurious
and consequently both sides of (\ref{res}) are
polynomials.
Note that
\begin{equation}
f_{\lambda }^{(n l)} (z_1 \tau ^3 |
z'|z'', z_1 \tau ^4 )
=(-\tau )^{l-n+\lambda -2}
(1-\tau ^2 )\varphi ^{(n~ l-2)}_{\lambda~ l-n+\lambda -2}
(z'|z'')
\label{f1}
\end{equation}
is independent of $z_1$.
Because of (\ref{blin}), (\ref{Atr}) and (\ref{f1})
the degree with respect to $z_1 $
of the RHS is $n-1$ and
that of LHS is at most $n-1$.
Hence (\ref{res}) is a polynomial equation
in $z_1 $ of degree $n-1$.
Owing to Claim 1,
$n$ points $z_1 =z_j \tau ^{-2}$
(~$2 \leq j \leq n+1$~)
satisfy (\ref{res}).
Thus it holds identically.

Therefore $Cyc (n,l)$ is verified. ~~~~$\Box$

{}~

The second equation (\ref{eqn:C}) is satisfied
if the following proposition is valid:
\begin{prop}
\begin{equation}
\begin{array}{rl}
&{\rm det} \left(
\overline{A}_{\lambda }^{(n l)}
(x_{\mu }| z_1 \tau ^{-4}, z_2 , \cdots , z_{n+1} |
z_{n+2} , \cdots, z_N )
\right) _{1\leq \lambda , \mu \leq n} \\
=&
(-\tau )^{2-l}
\overline{\Delta }^{(n l)}(x_1 , \cdots , x_n |
                        z_2 , \cdots , z_{n} |
                        z_{n+1}, \cdots , z_N |z_1 ),
\end{array}
\label{eDAC}
\end{equation}
where
$$
\begin{array}{cl}
&\displaystyle\overline{A}_{\lambda }^{(n l)}
(x|z_1 \tau ^{-4},z_2 , \cdots , z_n
| z_{n+1}, \cdots , z_N ) \\
=&
\displaystyle\frac{x-z_1 \tau ^{-1}}{x-z_1 \tau ^{-3}}
A_{\lambda }^{(n l)} (x|
z_1 \tau ^{-4}, z_2 , \cdots , z_n
| z_{n+1}, \cdots , z_N ) \\
+&
\displaystyle\frac{z_1 (-\tau )^{N-4}}{x}
\left\{
\frac{x-z_1 \tau ^{-1}}{x-z_1 \tau ^{-3}}
\prod_{j=2}^{N} \frac{x-z_j \tau ^{-1}}
{z_j -z_1 \tau ^{-2}}
-\tau ^{-2(N-2)}
\prod_{j=2}^{N}
\frac{x-z_j \tau }{z_j -z_1 \tau ^{-2}}
\right\} \times \\
\times & A_{\lambda }^{(n l)} (z_1 \tau ^{-3}|
 z_1 \tau ^{-4}, z_2 , \cdots , z_n |z_{n+1}, \cdots , z_N ),
\end{array}
$$
and
\begin{eqnarray}
&&\overline{\Delta }^{(n l)}(x_1 , \cdots , x_n |
                        z_2 , \cdots , z_{n} |
                        z_{n+1}, \cdots , z_N |z_1 ) \nonumber \\
& = & \displaystyle\prod_{j=n+1}^{N}
 \frac{z_j -z_1 \tau ^{-2}}{(z_1 -z_j )\tau ^{-1}}
\Delta ^{(n l)}
(x_1 , \cdots , x_n | z_1 , \cdots , z_{n} |
                      z_{n+1}, \cdots , z_N )
\nonumber \\
& - & \displaystyle\sum_{j=n+1}^N
 \frac{(1-\tau ^{-2})z_j }{(z_1 -z_j )\tau ^{-1}}
 \prod_{k=n+1 \atop k\neq j}^{N}
 \frac{z_k -z_1 \tau ^{-2}}{(z_j -z_k )\tau ^{-1}}
 \prod_{i=2}^{n}
 \frac{z_i -z_j \tau ^{-2}}{z_i -z_1 \tau ^{-2}}
\Delta ^{(n l)}
(x_1 , \cdots , x_n |
z_2 , \cdots , z_{n}, z_j |
z_{n+1}, \stackrel{j}{\hat{\cdots }}, z_N , z_1 ). \nonumber
\end{eqnarray}
\label{DAC}
\end{prop}
{\sl Proof.}~~~~In a similar way as in
Proposition \ref{main}
the following two claims can be shown:

Claim $1'$.~~~~When
$z_j =z_1 \tau ^{-2}$ ($n+1 \leq j \leq N$ )
both sides of (\ref{eDAC}) coincide.

Claim $2'$.~~~~When
$z_i =z_j \tau ^{-2}$ ($2\leq i \leq n < j \leq N$)
both sides of (\ref{eDAC}) coincide.

{}From power counting we do not need an analog of Claim 3.
In fact, let us factor out the difference product
$\displaystyle\prod_{\mu < \nu } (x_{\mu }-x_{\nu })$
and
$\displaystyle\prod_{j=2}^{n+1} \left(
(z_j -z_1 \tau ^{-2} ) \prod_{i=n+2}^{N} (z_i -z_j \tau ^{-2} )
\right) $
from the difference of both sides.
Then it is a rational function
of homogeneous degree $-n$ with at most
$n-1$ simple poles located
at $z_i =z_1 \tau ^{-2}$ ($2\leq i \leq n$)
and therefore it should be zero. ~~~~$\Box$

{}~

Now we are in a position to describe
the main theorem of the present paper.

\begin{thm}~~~~For $n<l$, the
integral formula given by
$$
H(z_1,\cdots,z_N)
=\prod_{\mu =1}^{n} \oint_{C} dx_{\mu }
F(x_1 , \cdots , x_n | z_1 , \cdots , z_N )
\Psi (x_1 , \cdots , x_n | z_1 , \cdots , z_N ),
$$
$$
F(x_1 , \cdots , x_n | z_1 , \cdots ,z_N )
=\frac{\Delta ^{(n l)}(x_1 , \cdots , x_n | z_1 , \cdots , z_n |
                      z_{n+1}, \cdots , z_N )}
      {\displaystyle\prod_{j=1}^{n}\prod_{i=n+1}^{N}
       (z_i -z_j \tau ^2 )},
$$
$$
\Delta ^{(n l)}(x_1 , \cdots , x_n | z_1 , \cdots , z_n |
                      z_{n+1}, \cdots , z_N )
=\det \left( A_{\lambda }^{(n l)} (x_{\mu }|
z_1 , \cdots , z_n |
                      z_{n+1}, \cdots , z_N )
\right) _{1\leq \lambda , \mu \leq n},
$$
$$
\begin{array}{rl}
A_{\lambda }^{(n l)} (x| a_1 , \cdots , a_n |
b_1 , \cdots b_l ):= &
\displaystyle\prod_{j=1}^{n} (x-a_j \tau )
f_{\lambda }^{(n l)} (x| a_1 , \cdots , a_n | b_1 , \cdots ,b_l ) \\
& + \tau ^{2(l-n+\lambda -1)}
\displaystyle\prod_{i=1}^{l} (x-b_i \tau ^{-1})
g_{\lambda }^{(n)} (x| a_1 , \cdots , a_n ),
\end{array}
$$
satisfies
(\ref{eqn:B}) and (\ref{eqn:C})
with $(\d_+ , \d_- )=(\tau ^{-n} , \tau ^{-l} )$.
\end{thm}

{}~

Finally we give Smirnov's formula for the case $n=l$.
The above formula specialized to $n=l$ does not give a
solution. We use an $(n-1)$-fold integration, and
replace $\Delta^{(nn)}$ by
\eq
\Delta^{(n n)}(x_1,\cdots,x_{n-1}|a_1,\cdots,a_n|b_1,\cdots,b_n)
=\hbox{\rm det\/}
\left(A^{(n n)}_{\lambda+1}(x_\mu|a_1,\cdots,a_n|b_1,\cdots,b_n)\right)
_{1\leq\lambda,\mu\leq n-1}.
\endeq
In this case
$f^{(n n)}_\lambda(x|a_1,\cdots,a_n|b_1,\cdots,b_n)
=g^{(n)}_\lambda(x|b_1,\cdots,b_n)$ and $g^{(n)}_1(x|b_1,\cdots,b_n)=0$,
and hence
\[
\begin{array}{rl}
A_{\lambda }^{(nn)} (x| a_1 , \cdots , a_n |
b_1 , \cdots b_n )= &
\displaystyle\prod_{j=1}^{n} (x-a_j \tau )
\,g_{\lambda }^{(n)} (x| b_1 , \cdots ,b_n ) \\
& + \tau ^{2(\lambda -1)}
\displaystyle\prod_{i=1}^{n} (x-b_i \tau ^{-1})
\,g_{\lambda }^{(n)} (x| a_1 , \cdots , a_n ).
\end{array}
\]

\section*{Acknowledgements}
We would like to thank A. Nakayashiki and F. A. Smirnov
for a number of discussions.
This work is partly supported by the Grant-in-Aid for
Scientific Research from the Ministry of Education,
Science and Culture, Japan,
Nos. 04245105, 05402001, 05452004  and 04-2297.

\end{document}